\documentclass[12pt]{article}
\usepackage{graphicx}
\usepackage{amssymb}

\newsavebox{\sboxpubnumber}
\newsavebox{\sboxpubdate}
\newcommand{\pubdate}[1]{\begin{lrbox}{\sboxpubdate}{#1}\end{lrbox}}

\newcommand{\Title}[1]{\begin{center} {\Large #1 } \end{center}}
\newcommand{\Author}[1]{\begin{center}{ \sc #1} \end{center}}
\newcommand{\Address}[1]{\begin{center}{ \it #1} \end{center}}

\newenvironment{Abstract}{\begin{quotation}  }{\end{quotation}}
\newenvironment{Presented}{\begin{quotation} \begin{center}
             PRESENTED AT\end{center}\bigskip
      \begin{center}\begin{large}}{\end{large}\end{center}
      \end{quotation}}
\newcommand{\Acknowledgements}{\bigskip  \bigskip \begin{center} \begin{large}
             \bf ACKNOWLEDGEMENTS \end{large}\end{center}}

\newcommand{\prd}{Phys.~Rev.~D}

\begin{document}

\begin{titlepage}
\pubdate{\today}                    

\vfill
\Title{Cosmological consequences of short distance physics}
\vfill
\Author{Jens C. Niemeyer}
\Address{Max-Planck-Institut f\"ur Astrophysik\\
Karl-Schwarzschild-Str. 1, D-85748 Garching, Germany}
\vfill
\begin{Abstract}
Inflation can act as a space-time microscope for Planck or string
scale effects, leaving potentially observable traces in the
primordial perturbation spectrum. I discuss two frameworks that were
used recently to study this phenomenon: nonlinear dispersion and short
distance uncertainty.
\end{Abstract}
\vfill
\begin{Presented}
    COSMO-01 \\
    Rovaniemi, Finland, \\
    August 29 -- September 4, 2001
\end{Presented}
\vfill
\end{titlepage}
\def\thefootnote{\fnsymbol{footnote}}
\setcounter{footnote}{0}

\section{Introduction}

To the best of our current understanding, inflation -- an epoch of
accelerated cosmological expansion -- may have taken place when the
expansion rate of the universe was $H \sim 10^{14}$ GeV (but
there exist models at lower values of $H$) \cite{L90,L01}. Quantum 
vacuum fluctuations at this energy scale are generally held responsible
for producing adiabatic perturbations that are reflected in the CMBR
anisotropies. It is possible that nature has provided us with a rare glimpse
at physics far beyond any other means of experimental access. 

The increasing precision of observations linked, directly or indirectly,
to the physics of inflation presents us with two important
challenges. First, we must probe the sensitivity of cosmological
predictions derived in the framework of semiclassical QFT
on curved or non-stationary space-times with respect to possible
quantum gravitational modifications. This approach has a long and
fruitful tradition in the context of the so-called trans-Planckian
problem of Hawking radiation (reviewed in \cite{J00}) and was indeed
inspired by it. Much of the work discussed in Sec.~\ref{sonic} is based
on fluid analogies of black hole horizons introduced by Unruh
\cite{U95} which study the effects of a nonlinear dispersion relation
on outgoing Hawking quanta. Since these are often described as ``sonic
black holes'' I will refer to this line of research as ``sonic
inflation'' in Sec.~\ref{sonic}. 

Secondly, and perhaps even more
importantly, it is possible that such modifications leave an
unambiguous and detectable footprint in the cosmological
data. Its concrete quantification could eventually be
based on string/M theory models (or directly on the holographic
principle \cite{H02}), but one can begin this ambitious task by exploring
field theoretic models that mimic some general
feature of short distance physics (as an analogy, think of starting
with the Euler equations and add the transport terms without knowing
the microscopic theory). For example, it is possible to
implement a cutoff that effectively limits the resolution of short
distances into the standard formalism, and re-analyze the predictions
of inflation. I discuss the current status of this program in
Sec.~\ref{sdu}. 

All of these studies break local Lorentz invariance in order to define
the notion of a locally measurable short distance
scale. So far, there is no experimental indication that Lorentz
invariance might be violated in nature \cite{LJM01}. On the other hand, neither 
is there any experimental indication that it is unbroken at  $\sim
10^{14}$ GeV, so one should remain open minded. As for physical
motivation, there are speculations about 
broken Lorentz invariance in the contexts of string/D-brane
interactions \cite{EMN00}, brane cosmology \cite{CEG01}, and 
quantum gravity phenomenology \cite{A01}, to name a few. Violations
or modifications of 
local Lorentz symmetry also play a vital role in varying speed of
light (VSL) cosmologies, making it obvious to look at VSL from the
viewpoint of short distance uncertainty; see Sec.~\ref{vsl} for a
brief discussion.

\section{Sonic inflation}
\label{sonic}

The first thing an experimentalist notices as the wavelength of a
fluid sound wave approaches the molecular scale is a change of the
dispersion relation, typically dipping off from the linear behavior
toward a slower growth of frequency $\omega$ with wavenumber $\kappa$. This
fact, together with the observation that sound waves in a fluid with
sonic horizon are mathematically equivalent to scalar fields in a
black hole space-time, led Unruh \cite{U95} to introduce sonic black holes
(or ``dumb holes'') by replacing the linear dispersion relation in the
mode equation of Hawking quanta with an artificial nonlinear one,
chosen to mimic the fluid behavior. Interpreting the scale of the
nonlinearity as the Planck scale, this framework has been used
extensively to study the robustness of Hawking radiation with respect
to trans-Planckian physics \cite{BMPe95}.

Similar to Hawking photons barely escaping to infinity, quantum
fluctuations in an inflationary background undergo enormous
redshifting before they reach super-horizon scales. Trans-Planckian
effects in inflation can therefore be addressed in a nearly identical
manner \cite{MB00,N00}. Both scalar and tensor perturbation amplitudes are
proportional to the solution of a harmonic oscillator equation, 
\begin{equation}
\label{mode}
\chi_k'' + \omega^2 \chi_k = 0\,\,,
\end{equation}
for the mode $\chi_k$ of the comoving wavenumber $k$ as a function of
conformal time $\eta$, evaluated at horizon crossing, $k = a H$ ($a$
is the scale factor of a flat FRW line element and
$H=a'/a^2$). Normally, $\omega^2$ is given by
\begin{equation}
\label{omega}
\omega_0^2(\eta) =  k^2 - \frac{a''(\eta)}{a(\eta)} \,,
\end{equation}
whereas it is written as 
\begin{equation}
\label{omegaF}
\omega_F^2(\eta) = [a(\eta) \,F(k/a(\eta))]^2 -
\frac{a''(\eta)}{a(\eta)} \,,
\end{equation}
in the modified form. $F(k/a)$ is a smooth function with the
following properties:
\begin{enumerate}
\item $F(k/a)$ converges to the usual  physical wavenumber $\kappa=
k/a$ for $\kappa \ll \kappa_c$.  The critical wavenumber $\kappa_c$
marks the scale of deviations from standard physics and is presumably
of order the Planck scale. It is usually assumed that $F(k/a) = \kappa$ at 
horizon crossing, but this need not be the case even if $H
< \kappa_c$, depending on the shape of $F$ \cite{N00}.
\item At around $\kappa_c$, $F$ veers off the linear path and
increases, becomes constant, or decreases in what are usually called
superluminous, Unruh, or subluminous dispersion relations \cite{CJ96}. 
\end{enumerate}
With the exception of \cite{KG00} where it is derived from
$\kappa$-Poincar\'e algebra, $F$ is simply chosen by hand to fit these
criteria. On a side note, it is even possible to create forms of $F$
that allow for an inflationary period without a scalar inflaton field
\cite{ABM01}. 

As there are two dimensionful variables in this theory, $H$ and
$\kappa_c$, the solutions are controlled by the dimensionless quantity  
\begin{equation}
\label{sigdef}
\sigma = \frac{H}{\kappa_c}
\end{equation}
which is usually a small number. Another important role is played by
the adiabaticity parameter,
\begin{equation}
\label{condition}
{\cal C}(\eta) = \left|\frac{\omega'}{\omega^2}\right|\,\,,
\end{equation}
describing the ``smoothness'' of $F$ with respect to the oscillations
of $\chi_k$ and hence the accuracy of the WKB approximation. 

As shown in \cite{NP01} and confirmed in \cite{S01}, ${\cal C} \leq
\sigma$ for all regular, monotonically growing functions
$F$, while changes in the power spectrum due to nonadiabatic particle
production are at most of order ${\cal C}$ (but may be much
smaller). Consequently, if the 
critical wavenumber is identified with the Planck scale, the CMBR
fluctuation amplitude indicates that $\sigma \lesssim 10^{-5}$ and such
changes are undetectable. However, in some string or M theory models
$\sigma$ may be several orders of magnitude larger \cite{Kea02}.

Singular dispersion relations of the form $F(\kappa) \sim
\kappa (1-\kappa/\kappa_c)^{-n}$ with $0 < n < 1$ and those with
$F(\kappa \gg \kappa_c) < H$ exhibit nonadiabatic behavior \cite{NP01}. In
the latter case, each mode undergoes another ``horizon crossing'' on
very small length scales below which it is frozen by cosmic
expansion. Ref.~\cite{MBP01} argues that this effect produces an
effective dark energy contribution for a certain class of dispersion
relations (but see \cite{Lea02} who reach a different conclusion based
on the analysis of the stress-energy tensor). A string theoretic
derivation of this type of dispersion relation is offered in
Ref.~\cite{BFM01}.  

In order to find the initial conditions of the mode equation
(\ref{mode}) one must choose a vacuum for the field. The adiabatic
vacuum, constructed from the WKB solution of Eq.~(\ref{mode}), can be
considered a natural ground state \cite{BD84}. Backreaction of Planck
scale particles forces the vacuum to be close to the adiabatic one in
order to be consistent with slow-roll inflation \cite{LL93}. In this case,
deviations of the power spectrum can be caused by nonadiabatic processes on
subhorizon scales (see above) or if $\omega_F \ne \omega_0$ at horizon
crossing. Alternative vacua were investigated in \cite{MB00}. The authors of
Ref.~\cite{BJM01} gave  the initial value problem a new twist by
constructing a bouncing universe cosmology. They followed 
the mode evolution through the bounce where it picks up
non-adiabatic corrections, showing that the spectrum may be affected
even with adiabatic initial conditions.

A crucial question is whether the effects of nonlinear dispersion
can be disentangled from features of the inflaton potential. As noted
in \cite{HK01} and further analyzed in \cite{Kea02}, Planck scale effects might
act differently on scalar and 
tensor perturbations and thereby violate the scalar-tensor
consistency relation of slow-roll inflation. This violation
could be the smoking gun for physics beyond the standard paradigm of
inflation.  

\section{Short distance uncertainty}
\label{sdu}

Instead of modifying the dispersion relation by hand one can use an
observation made in various contexts in string theory and quantum
gravity: on very general grounds, our experimental ability to probe
short distances seems to be limited by the Planck or string scale
\cite{GM88}. The cutoff presumably arises from
complicated dynamics of the 
underlying fundamental theory but it can also be modeled by a nonlinear
correction to the canonical commutation relation:
\begin{equation}
\label{commutation}
[{\bf x},{\bf p}]= i \hbar (1 + \beta {\bf p}^2)
\end{equation}
as discussed in \cite{K94}. The correction term results in a lower
bound $\Delta x_{\rm min} \sim \beta^{1/2}$ for distance
measurements. Furthermore, Eq.~(\ref{commutation})
belongs to a class of only very few types of short-distance structures
of space-time that are admitted under very general assumptions
\cite{K98}. Hilbert space representations of Eq.~\ref{commutation} were
employed, for instance, for regularizing field theory \cite{KM97} and, more
recently, for analyzing the impact of short distance uncertainty on
the predictions of inflation.

In Ref.~\cite{K00}, Eq.~(\ref{commutation}) was implemented in a
scalar field theory (representing, e.g., the inflaton) on an FRW
background. The action decomposes into modes of conserved quasi-wavenumber
$\tilde k$ which converges to the usual comoving wavenumber $k$ for
$\tilde k/a \ll \beta^{-1/2}$. Each $\tilde k$-mode is generated at
the time $\eta_c$ corresponding to its ``Planck scale crossing'',
defined by $a(\eta_c) = \tilde k (e \beta)^{1/2}$, and obeys an
oscillator equation with mass and damping terms that are singular at
$\eta_c$. 

Refs.~\cite{KN01,Eea01} analyzed the implications for the cosmic
perturbation spectrum. As shown in Ref.~\cite{KN01} by transforming
the $\tilde k$-mode  equation to the form of Eq.~(\ref{mode}),  new effects 
assuming full adiabaticity shortly before horizon crossing are at most of order
$\sigma^2$. However,  the numerical evaluation of ${\cal C}$ indicates
that ${\cal C} \sim \sigma$ during most of the  
subhorizon evolution so that, following the arguments in
Sec.~\ref{sonic}, O($\sigma$) effects on the spectrum cannot be strictly ruled
out even if the mode is adiabatic immediately after its creation.

A numerical analysis of the $\tilde k$-mode evolution was performed in
Refs.~\cite{Eea01}. The authors used initial conditions obtained from
an analytical solution of the linearized mode equation near $\eta_c$
(also proposed in \cite{KN01}) and found new features in the power spectrum
whose amplitudes
are linear in $\sigma$. These are apparently related to deviations from
adiabaticity. Ref.~\cite{Kea02}, on the other hand, argues on the
basis of low energy locality that the leading order effect must be
O($\sigma^2$). 

Obviously, finding the leading order correction in $\sigma$ is of
utmost importance for the observability of cutoff effects in
the CMBR, given that $\sigma$ is probably $\lesssim 10^{-3}$. A linear
signal would make the detection of Planck scale physics in
cosmological data a realistic goal for the intermediate future. But
even O($\sigma^2$) effects are potentially 
observable in certain classes of M theoretic models \cite{Kea02}. Much of
the remaining uncertainty stems from the ambiguity in the initial
conditions of the $\tilde k$-mode equation. However, in contrast with
the sonic inflation approach, we now have a specific prescription for the
generation of individual $\tilde k$-modes, so there is hope that the model
itself picks out a preferred vacuum. For completeness, note also that
the generation of modes from a quantum gravitational ``soup'' plays a
central role in Brout's two-fluid model of inflation \cite{B01}.

\section{Varying speed of light cosmology}
\label{vsl}

Varying speed of light (VSL) cosmology is one of the few serious competitors
of inflation \cite{M93a}. It solves the horizon problem by postulating an
early cosmological epoch where, roughly speaking, information
propagates faster than the current speed of light $c$. This involves
some modification or violation of Lorentz symmetry whose form and
implementation has been addressed in different ways (e.g.,
\cite{Bea00}). A number of 
possible explanations for a varying speed of light have been offered,
motivated, e.g., by noncritical string theory \cite{EMN93}, brane
cosmology \cite{K99}, and non-commutative geometry \cite{AM01}. 

The framework of short distance uncertainty described in
Sec.~\ref{sdu} provides a well-defined platform for searching for VSL
effects. As shown in Ref.~\cite{N01} (see also \cite{L00}), the reduction of
the available phase space volume 
per quantum mode at short wavelengths causes the equation of state of
ultrarelativistic particles to stiffen at very high densities. This
leads to a stronger than usual deceleration of the scale factor
which competes with a higher than usual propagation speed of the
particles. The definition of the latter, however, is ambiguous:
possibilities include the
group and phase velocity in the high energy tail, the thermal average
of the group and phase
velocity, and the speed of sound. Of these three groups, only the
first provides a possible solution to the cosmological horizon
problem. Perhaps more worrisome is the fact that this solution
involves applying Eq.~(\ref{commutation}) at super-Planckian densities
where it is unlikely to fully represent the relevant physics.

To conclude, the field of short distance physics in the very early
universe has opened a number of new ways to think about old problems,
including the tantalizing (but still highly speculative) prospect of
detecting Planck scale effects in cosmological observations. We can
hope for interesting new developments by the time of Cosmo-02.

\Acknowledgements
I thank Achim Kempf for helpful comments and suggestions.

\end{document}